\documentclass[twocolumn,showpacs,preprintnumbers,amsmath,amssymb,superscriptaddress]{revtex4}

\usepackage{graphicx}
\usepackage{dcolumn}
\usepackage{amsmath}    
\usepackage{verbatim}   
\usepackage{color}      
\usepackage{subfigure}  
\usepackage{hyperref}   
\usepackage{bm}

\begin{document}

 \title{
Voltage tunability of single spin-states in a quantum dot

}

\author{Anthony J. Bennett}
\email{anthony.bennett@crl.toshiba.co.uk}
\thanks{Corresponding author}
\affiliation{Toshiba Research Europe Limited, Cambridge Research Laboratory,\\
208 Science Park, Milton Road, Cambridge, CB4 0GZ, U. K.}

\author{Matthew A. Pooley}
\affiliation{Toshiba Research Europe Limited, Cambridge Research Laboratory,\\
208 Science Park, Milton Road, Cambridge, CB4 0GZ, U. K.}
\affiliation{Cavendish Laboratory, Cambridge University,\\
J. J. Thomson Avenue, Cambridge, CB3 0HE, U. K.}

\author{Yameng Cao}
\affiliation{Toshiba Research Europe Limited, Cambridge Research Laboratory,\\
208 Science Park, Milton Road, Cambridge, CB4 0GZ, U. K.}
\affiliation{Department of Physics, Imperial College London, Prince
Consort Road, London SW7 2AZ, U. K. }

\author{Niklas Sk\"{o}ld}
\affiliation{Toshiba Research Europe Limited, Cambridge Research Laboratory,\\
208 Science Park, Milton Road, Cambridge, CB4 0GZ, U. K.}

\author{Ian Farrer}
\affiliation{Cavendish Laboratory, Cambridge University,\\
J. J. Thomson Avenue, Cambridge, CB3 0HE, U. K.}

\author{David A. Ritchie}
\affiliation{Cavendish Laboratory, Cambridge University,\\
J. J. Thomson Avenue, Cambridge, CB3 0HE, U. K.}

\author{Andrew J. Shields}
\affiliation{Toshiba Research Europe Limited, Cambridge Research Laboratory,\\
208 Science Park, Milton Road, Cambridge, CB4 0GZ, U. K.}

\date{\today}%




\begin{abstract}

Single spins in the solid-state offer a unique opportunity to store
and manipulate quantum information, and to perform quantum-enhanced
sensing of local fields and charges. Optical control of these
systems using techniques developed in atomic physics has yet to
exploit all the advantages of the solid-state. We demonstrate
voltage tunability of the spin energy levels in a single quantum dot
by modifying how spins sense magnetic field. We find the in-plane
$g$-factor varies discontinuously for electrons, as more holes are
loaded onto the dot. In contrast, the in-plane hole $g$-factor
varies continuously. The device can change the sign of the in-plane
$g$-factor of a single hole, at which point an avoided crossing is
observed in the two spin eigenstates. This is exactly what is
required for universal control of a single spin with a single
electrical gate.

\end{abstract}

\maketitle 

The spin of charges in quantum dots (QDs) has long been considered a
suitable qubit for quantum operations \cite{Loss98}. The three
dimensional confinement offered by a single semiconductor QD reduces
many decoherence mechanisms, allowing impressively long coherence
times to be observed in coherent population trapping
\cite{Brunner09} or using spin-echo techniques \cite{Press10,
DeGreve2011}. In the latter, control of the spins was achieved with
resonant, ultrafast optical pulses. An alternative mechanism for
controlling single spins is for an electric field to vary their
coupling to a fixed magnetic field (\textbf{\emph{B}}), described by
the $g$-tensor (\textbf{\emph{g}}) \cite{Pingenot08}. This method
allows multiple closely spaced spin qubits to be individually
addressed via nano-electrodes, without resonant lasers or localised
magnetic fields. Critical to this concept is the ability to change
the sign of one component of the $g$-tensor \cite{Pingenot08,
Pingenot11}. Then through careful alignment of the magnetic field
direction it is possible to switch between two electric fields where
the precession directions of the spin (given by \textbf{\emph{g.B}})
are orthogonal on the Bloch sphere. In such a system `universal'
control can map any point on the Bloch sphere onto any other point.
Although experimental studies have been made of the $g$-tensor in
QDs \cite{Bayer1999B, Schwan11, Jovanov2011, Goddenarxiv12,
Nakaoka07, Deacon11}, the change of sign of one component with
electrical field has yet to reported.

Early work used semiconductor quantum wells to electronically tune
the $g$-tensor of multiple spins by shifting their carrier
wavefunctions into areas of different material composition
\cite{Salis01}. Extending this work to single charges trapped in
zero-dimensional structures has not been straightforward as the
carriers tunnel out of the structure when electric field is applied.

One approach was demonstrated using electronically coupled pairs of
QDs \cite{Doty06}. Carriers displayed the $g$-tensor of the material
in which they were located, so when a voltage was applied to
localise the charge in one dot the $g$-factor measured was that of
the indium-rich QD. However, when the wavefunction was delocalised
between the dots there was a much greater spatial overlap with AlAs
semiconductor in the barrier, and a change in $g$ was observed.

Recently, experiments showed that vertical electric fields can
change the $g$-factor relevant for out-of-plane magnetic fields
($g^{\perp}$) \cite{Jovanov2011} in dots that are engineered to have
increased height and reduced indium-composition. The in-plane
$g$-factor of an $s$-shell hole ($g^{\parallel}_{\text{h,s}}$) was
also modified by vertical electric field \cite{Goddenarxiv12} over a
modest field range of 20 kV/cm.  However, both of these measurements
were made in the photo-current regime, where carriers rapidly tunnel
from the dot greatly limiting the spin lifetime. Conversely, an
in-plane electric field can change the $g$-tensor but there the
tunneling problem is even more severe \cite{Nakaoka07, Deacon11}.

We solve these problems by locating single dots in the center of a
$p-i-n$ diode where barriers that hinder tunneling allow us to apply
electric fields, $F$, up to -500 kV/cm, whilst still observing
photoluminescence \cite{Bennett10}. We study changes in the
$g$-tensor as a function of electric field and show a high degree of
control can be achieved for both electrons and holes. We observe
that continuous variation in the $g$-factor of holes in a parallel
magnetic field can be obtained. Different behavior is observed
depending on whether the hole is in the $s$- or $p$-shell. When
$g^{\parallel}_{\text{h,s}}$ has a low value at zero electric field
these devices are capable of tuning it through an avoided crossing
at finite field, and changing its sign, without carriers escaping.

\begin{figure}[h]
\includegraphics[width = 90mm]{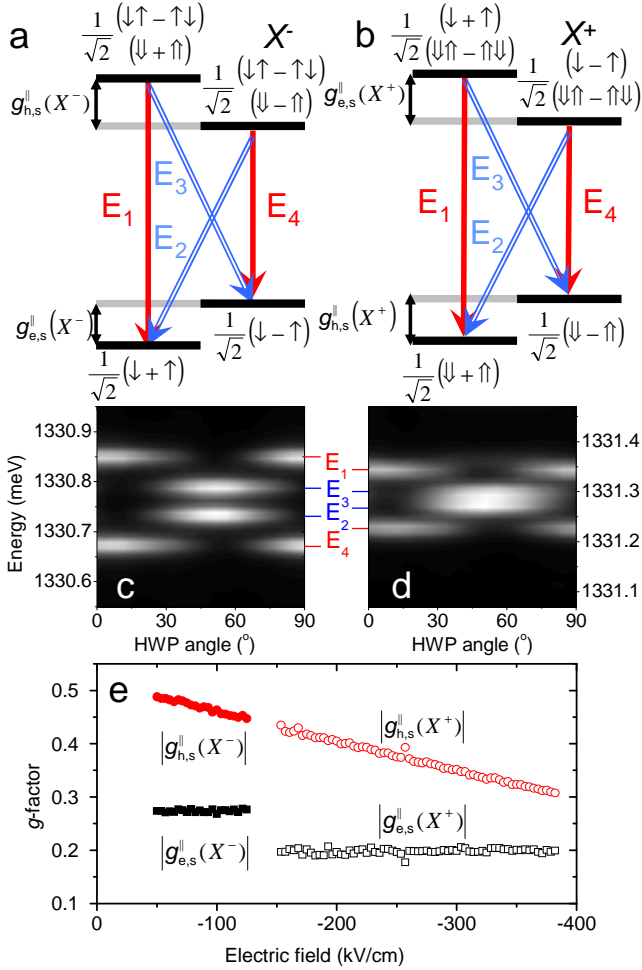} 
\caption{\label{Fig1} \textbf{ Electric field tuning of $s$-shell
electron and hole $g$-factors} (a) and (b) show the energy levels of
the negatively and positively charged excitons ($X^{-}$ and $X^{+}$,
respectively) in a Voigt geometry magnetic field. Transitions
$E_{1}$ and $E_{4}$ (red) result in linearly polarised emission
orthogonal to the magnetic field and $E_{2}$ and $E_{3}$ (blue) are
parallel to the magnetic field. (c) and (d) show polarisation
dependent spectra from the $X^{-}$ at -78.5 kV/cm and $X^{+}$ at
-385.7 kV/cm (respectively) at a field of 4T, as a half wave plate
(HWP) is rotated. (e) shows the extracted $s$-shell electron
$g$-factor, $|g^{\parallel}_{e,s}|$, and $s$-shell hole $g$-factor,
$|g^{\parallel}_{\text{h,s}}|$, as a function of electric field.}
\end{figure}

\parskip 5mm

\textbf{Results}

\parskip 5mm

\textbf{Charged exciton transitions in magnetic and electric field}.
When the magnetic field is orthogonal to the plane of the sample
(Faraday geometry, $B^{\perp}$) we see that the $g$-factors are
barely affected by electric field (see Supplementary Figure S1 and
Supplementary Notes). However, when the magnetic field is aligned in
the plane of the sample (Voigt geometry, $B^{\parallel}$) strikingly
different behavior is observed. The separate in-plane $g$-factors of
an $s$-shell electrons ($g^{\parallel}_{e,s}$) and holes
($g^{\parallel}_{\text{h,s}}$) may be determined from the decay
energies of the positively ($X^{+}$) and negatively ($X^{-}$)
charged excitons. The magnetic field splits both the upper
($g^{\parallel}_{\text{h,s}}\mu_{\text{B}} B^{\parallel}$) and lower
states ($g^{\parallel}_{\text{e,s}}\mu_{\text{B}} B^{\parallel}$) of
$X^{-}$, where $\mu_{\text{B}}$ is the Bohr Magneton. Four
transitions (($E_{1}$ to $E_{4}$)) are observed as shown in Figure
\ref{Fig1}a,b. The highest and lowest energy transitions of this
quadruplet ($E_{1}$ and $E_{4}$) emit photons with electric field
orthogonal to \textbf{\emph{B}} and the intermediate transitions
($E_{2}$ and $E_{3}$) parallel to \textbf{\emph{B}}. Fitting the
energies of each transition resulting from the $X^{-}$ state one can
determine $g_{e,s}^{\parallel} (X^{-})$ and $g_{h,s}^{\parallel}
(X^{-})$ using $g_{e,s}^{\parallel}(X^{-}) \mu_{\text{B}}
B^{\parallel} = E_{1}-E_{3} = E_{2}-E_{4}$ and
$g_{h,s}^{\parallel}(X^{-}) \mu_{\text{B}} B^{\parallel} =
E_{1}-E_{2} = E_{3}-E_{4}$. Similar arguments can be made to
determine $g_{e,s}^{\parallel}(X^{+})$ and
$g_{h,s}^{\parallel}(X^{+})$ from the $X^{+}$ transitions. There is
not enough information in this measurement alone to determine the
sign of these $g$-factors. However, for nearly all dots we see an
increase in the fine-structure splitting of the neutral exciton
state with magnetic field, which is a signature that both have the
same sign \cite{Stevenson06}, which we take to be negative
\cite{Doty06, Xu2007}.

$g^{\parallel}_{\text{e,s}}$ appears to be constant for a given
exciton complex, but on switching between the $X^{+}$ and $X^{-}$
transitions an abrupt step is always observed. The reason for this
is that $g^{\parallel}_{\text{e,s}} (X^{+})$ is determined by the
initial state (when there are two holes also present in the dot).
These holes are better confined than the electron and provide a
coulomb attraction that reduces the extent of the electron
wavefunction, pushing $g^{\parallel}_{\text{e,s}}$ closer to $+$2
\cite{Pryor}. However, $g^{\parallel}_{\text{e,s}}$ when no holes
are present is determined from the final state of the $X^{-}$
transition. For the sample of 15 dots studied
$|g^{\parallel}_{\text{e,s}} (X^{-})|=0.266 \pm 0.012$ and
$|g^{\parallel}_{\text{e,s}}(X^{+})|=0.178 \pm 0.033$, where the
numbers quoted are the mean $\pm$ the standard deviation.

In contrast to the behavior of the electron,
$g^{\parallel}_{\text{h,s}}$ varies linearly with electric field for
both $X^{+}$ and $X^{-}$ in Fig. \ref{Fig1}e. We estimate the
discontinuity in $g^{\parallel}_{\text{h,s}}$ on switching between
$X^{+}$ and $X^{-}$ is on average an order of magnitude smaller than
the similar effect for $g^{\parallel}_{\text{e,s}}$, as expected
given the greater spatial extent of the electron wavefunction. There
is remarkable homogeneity in the rate at which
$g^{\parallel}_{\text{h,s}}$ can be tuned with electric field for
different dots, $\xi = |dg^{\parallel}_{\text{h,s}}/dF| = (5.7 \pm
1.5) \times 10^{-4}$ cm/kV where $|g^{\parallel}_{\text{h,s}}|$ at
zero electric field is $0.469 \pm 0.110$. The scatter in the value
of $g^{\parallel}_{\text{h,s}}$ at $F = 0$ is greater than the
comparable value for the electron, as this is affected more strongly
by variations in dot height and lateral size \cite{Pryor}. The rate
$\xi$ compares well with the recent publication of Godden {\it et
al} \cite{Goddenarxiv12} which determines
$|g^{\parallel}_{\text{h,s}}|$ from the energy splitting of the
$X^{-}$ in a time-resolved photo-current measurement. This paper
reports a linear variation in $g^{\parallel}_{\text{h,s}}$ at a rate
of $3.5 \times 10^{-4}$ cm/kV over a range of only 20 kV/cm. The
rate of shift is also of the same order of magnitude as that
predicted theoretically, for dots of greater height and uniform
composition\cite{Pingenot11}. It will be interesting to see whether
further theoretical work can fully explain the variations in the
$g$-tensor we observe.

\begin{figure*}[ht]
\includegraphics[width=180mm]{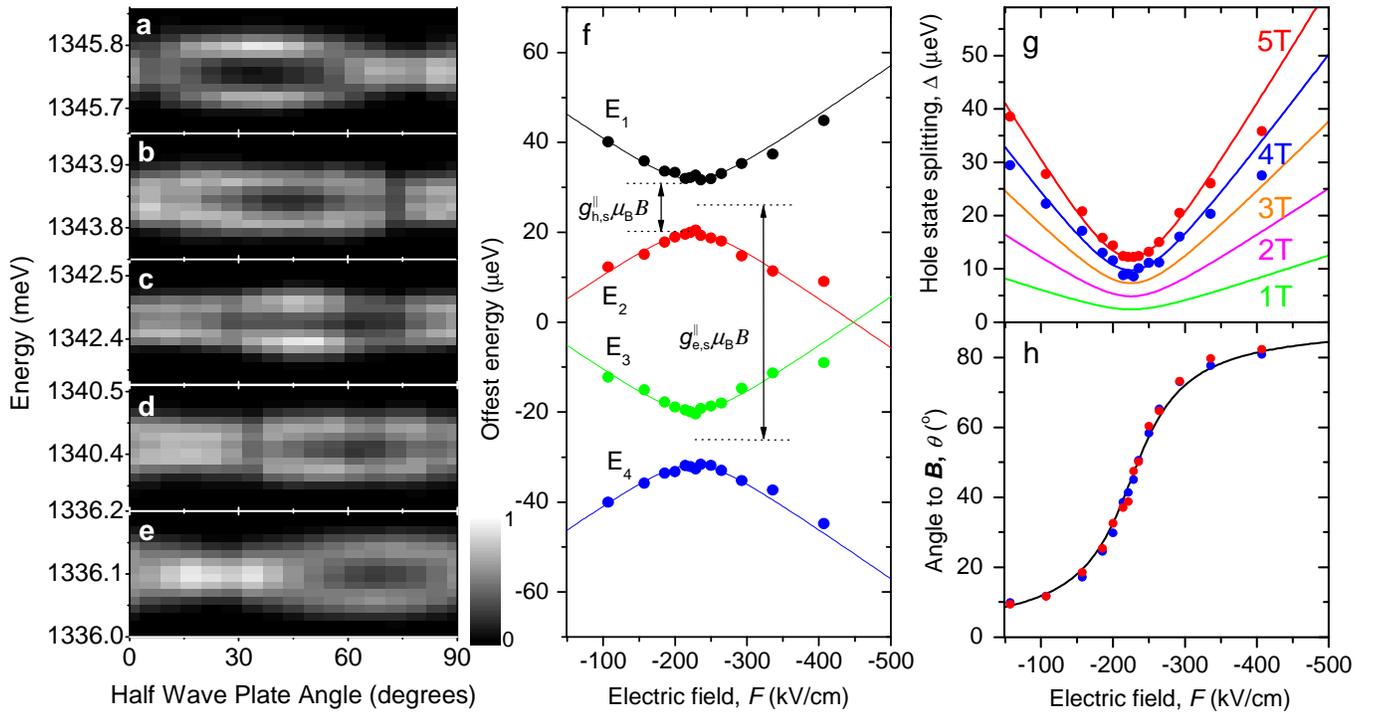}
\caption{\label{Fig2} \textbf{Changing the sign of the
Voigt-geometry $s$-shell hole $g$-factor with electric field}.
Polarized spectra of the positively-charged exciton ,$X^{+}$, for an
in-plane magnetic field of 5T and electric fields of (a) -107.1
kV/cm (b) -185.7 kV/cm (c) -221.4 kV/cm (d) -264.3 kV/cm and (e)
-335.7 kV/cm. (f) shows the energies of the four transitions of the
$X^{+}$ offset by their mean value at each electric field. (g) the
magnitude of the hole $g$-factor, $|g^{\parallel}_{\text{h,s}}|$ and
the (h) orientation ($\theta$) of the states relative to the
magnetic field. Both (g) and (h) show fits for 1 to 5T based on
equations \ref{Equ1} and \ref{Equ2}}.
\end{figure*}

\textbf{Minimising the $g$-factor of the $s$-shell hole.} With the
range of fields accessible in these samples, any dot with
$|g^{\parallel}_{\text{h,s}}| \leq$ 0.285 at $F=0$ can be tuned to a
minimum $|g^{\parallel}_{\text{h,s}}|$ at some field, $F_{0}$. We
now discuss data from a dot with $|g^{\parallel}_{\text{h,s}}| =$
0.174 at $F=0$. This same QD also displays a minimum in the
fine-structure splitting of the neutral exciton of 1.8 $\mu$eV at
-57.1 kV/cm. We find that $|dg^{\parallel}_{\text{h,s}}/dF| = 7.74
\times 10^{-4}$ cm/kV, and thus we are able to tune the hole
eigenstate splitting $\Delta =$ $|g^{\parallel}_{\text{h,s}}| \mu_B
B$ towards a minimum value at an electric field of $F_{0} =$ -225.0
kV/cm. For fields above $F_{0}$ (such as shown in Fig. \ref{Fig2}a)
we observe that the sign of $g^{\parallel}_{\text{h,s}}$ is the same
as for all other dots in the ensemble.  For electric fields below
$F_{0}$ (such as in Fig. \ref{Fig2}e) $g^{\parallel}_{\text{h,s}}$
has the opposite sign, which manifests itself as a clear difference
in the orientation angle at which the largest difference in
transition energies is observed. Fig. \ref{Fig2}f plots the $X^{+}$
transition energies as a function of electric field, $F$, to clearly
show the form of the anti-crossing in the hole states (the mean
value of all four transition energies has been subtracted for
clarity, to remove the Stark shift). The minimum hole state
splitting corresponds to $|g^{\parallel}_{\text{h,s}}(F_{0})|=$
0.042, but we stress that on either side of this minimum value the
$g^{\parallel}_{\text{h,s}}$ has different sign.

The behavior of $\Delta$ is reminiscent of the anti-crossing in
neutral exciton states that has been observed with electric field
\cite{Bennett10, Ghali2012, Trotta12}, however in this case the
states that are coupled together contain only a single hole. Indeed,
in the analysis of Plumhof $et$ $al$ \cite{Plumhof11}, who studied
the anti-crossing of the neutral exciton states under externally
applied strain, it was the hole wavefunction that dominated the
orientation of the eigenstates relative to the laboratory ($\theta$)
and anti-crossing of the eigenenergies. As with the neutral exciton
we fit the avoided crossing with a coupling parameter
$g^{\parallel}_{\text{h,s}}(F_{0}) \mu_{\text{B}} B^{\parallel}$,
where the splitting between the hole states varies linearly away
from $F_{0}$ at a rate $\xi \mu_{\text{B}} B^{\parallel}$.
\\
\linebreak
\begin{equation}
\label{Equ1}
 \Delta = g^{\parallel}_{\text{h,s}}(F) \mu_{\text{B}} B^{\parallel} = \mu_{\text{B}}
 B^{\parallel}
\sqrt{\xi^{2}(F-F_{0})^{2}+(g^{\parallel}_{\text{h,s}}(F_{0}))^{2}}
\end{equation}
\\
\linebreak
\begin{equation}
\label{Equ2} \theta = \pm
\text{tan}^{-1}\left[\frac{g^{\parallel}_{\text{h,s}}(F_{0})}{\xi
(F-F_{0}) \pm g^{\parallel}_{\text{h,s}}(F)}\right]
\end{equation}
\\

In Fig. \ref{Fig2}g and h we show data summarising the behavior of
the two hole eigenstates at 4T (the lowest field at which we can
spectrally resolve all four transitions) and 5T (the highest field
available with our magnet) fitted with this model. We observe that
the magnitude of the anti-crossing in energy appears to scale
linearly with $B^{\parallel}$, thus
$|g^{\parallel}_{\text{h,s}}(F_{0})|$ is constant, at least in the
range of fields we can probe. The resulting variation of $\theta$
with $F$ is the same for both magnetic fields (Fig. \ref{Fig2}h), in
accordance with equations \ref{Equ1} and \ref{Equ2}. It will be
interesting to further probe the behavior of this effect in higher
magnetic fields.

\begin{figure*}[ht]
\includegraphics[width=180mm]{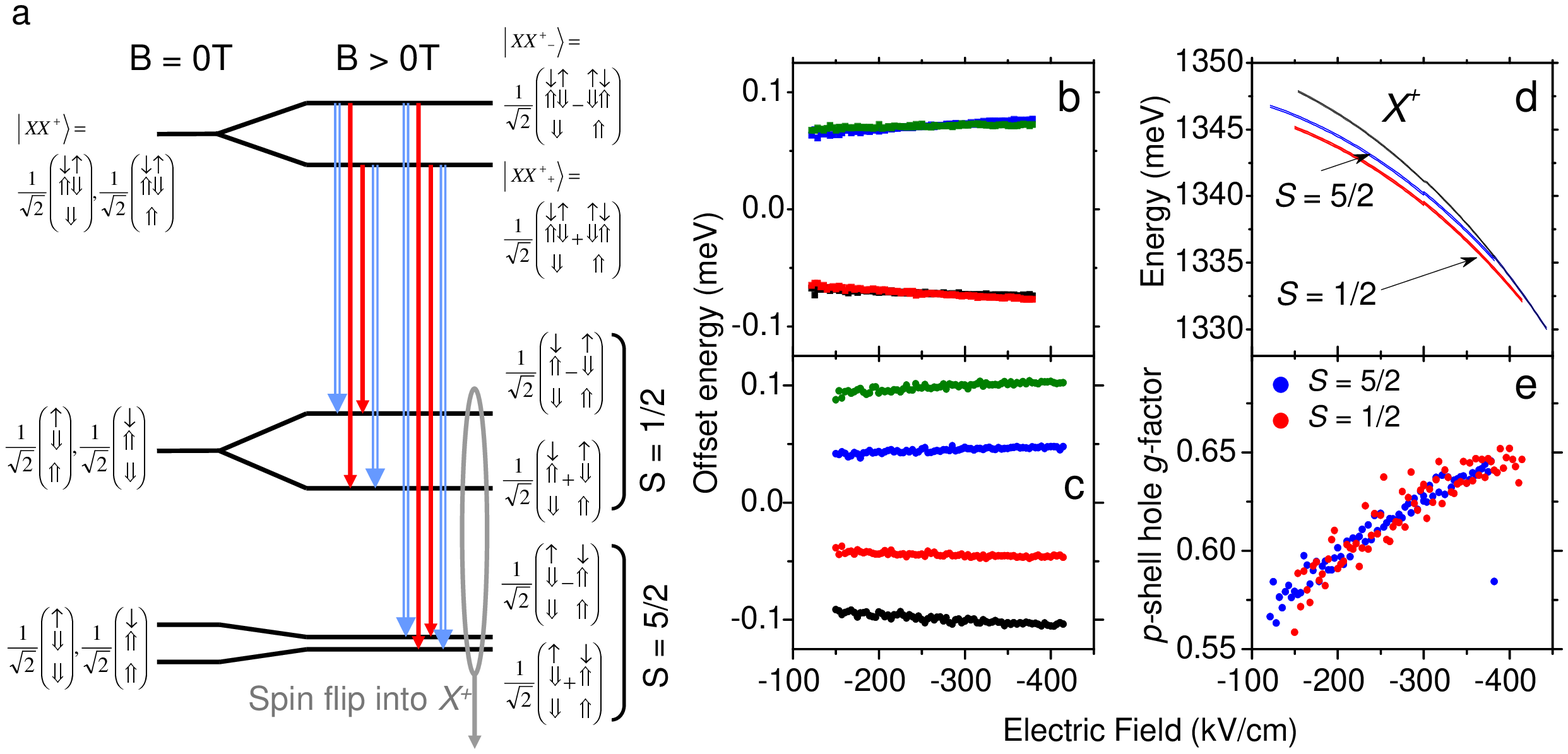}
\caption{\label{Fig3} \textbf{Determination of Voigt-geometry
$p$-shell hole $g$-factor}. (a) shows the allowed transitions for
recombination of an $s$-shell electron-hole pair of the
positively-charge biexciton, $XX^{+}$. Red and blue arrows indicate
photon emission with opposite linear polarisation. (b) the energies
of the quadruplet with $S =$ 5/2 final state, offset by their mean
at 4T, and (c) the energies of the quadruplet with $S =$ 1/2 final
state, offset by their mean at 4T, as a function of electric field.
(d) shows the absolute energy of the transitions shown in (a) versus
electric field, at 4T. From (b) and (c) we independently extract the
magnitude of the $p$-shell hole $g$-factor (e) for the $S =$ 1/2
(red) and $S =$ 5/2 (blue) transitions.}
\end{figure*}

\textbf{$g$-factor of the $p$-shell hole.} Finally, we study the
decay of the positively changed biexciton, $XX^{+}$, which consists
of a filled $s$-shell and a excess $p$-shell hole. These transitions
are observed on the low energy side of the $X^{+}$ transition
\cite{Rodt2005, Akimov2002}. We determine the $g$-factors of the
$p$-shell hole which, to our knowledge, has not been possible before
(although the Voigt geometry electron $p$-shell $g$-factor has been
probed \cite{Mayer06}). We find that in the Faraday geometry there
is no variation in the $p$-shell hole $g$-factor as a function of
electric field. In a Voigt geometry the brightest radiative decays
from $XX^{+}$ involve recombination of an $s$-shell electron and
hole. The resulting photons are linearly polarised as shown in Fig.
\ref{Fig3}a (blue and red arrows have orthogonal linear
polarisation) with the initial state $XX^{+}$ split by
$g^{\parallel}_{\text{h,p}} \mu_{\text{B}} B^{\parallel}$, where
$g^{\parallel}_{\text{h,p}}$ is the Voigt $p$-shell hole $g$-factor.
However, the final states can either have spin $S = $ 1/2 or 5/2:
their splittings are partly determined by the electron-hole exchange
between the $s$-shell electron and $p$-shell hole which has not been
well studied. Empirically, we see that the spin splitting of the $S
=$ 1/2 final state is below the system resolution at
$B^{\parallel}=$ 0, but increases with magnetic field. In contrast,
the $S =$ 5/2 final state has a spin splitting of several hundred
$\mu$eV at $B^{\parallel}=$ 0 but is reduced with $B^{\parallel}$.
Nevertheless, it is possible to measure the initial state splitting
$XX^{+}$, using either the $S =$ 1/2 or $S =$ 5/2 final state
quadruplets, and thus infer $|g^{\parallel}_{\text{h,p}}|$. When
this is done, both quadruplets lead to the same value of
$|g^{\parallel}_{\text{h,p}}|$ (Fig. \ref{Fig3}e), as expected. We
find that $|g^{\parallel}_{\text{h,p}}|$ has a greater magnitude
than $|g^{\parallel}_{\text{h,s}}|$ and varies non-linearly with
electric field. The greater extent of the $p$-shell hole
wavefunction outside the dot is likely to bring
$g^{\parallel}_{\text{h,p}}$ closer to the value determined by the
wetting layer and surrounding GaAs.

\parskip 5mm

\textbf{Discussion}

\parskip 5mm

Several proposals exist for universal control of a single spin in a
QD \cite{Loss98, Pingenot08, Pingenot11, Andlauer09}. The ability of
the device reported here to change the sign of
$|g^{\parallel}_{\text{h,s}}|$, combined with the reduced
hole-hyperfine interaction and greater hole spin lifetime open up
the possibility of all-electrical $4\pi$ manipulation of the hole
spin. Alternatively, controlled phase shifts may be achieved on a
qubit encoded on the spin of the electron by addition of two holes
for a predetermined time, which could be achieved by controlled
charging.

The timescale of any electrical control sequence is limited by the
resistance and capacitance of the diode to tens of picoseconds
\cite{Bennett08}, which is significantly greater than that achieved
with coherent optical pulses. However, the ability to achieve full
Bloch-sphere control with only a single electrical gate is a
promising avenue of investigation. Such a device could find
applications in a spin-based quantum memory \cite{Heiss10},
spin-echo techniques \cite{Press10, DeGreve2011} spin based quantum
computing \cite{Loss98, Folleti2009} and generation of photonic
cluster-states \cite{Lindner09}.

\parskip 5mm

\textbf{Methods}

\parskip 5mm

\textbf{Sample design.} The sample consists of a single layer of
self assembled quantum dots grown in the center of a 10nm wide GaAs
quantum well, clad with a 75$\%$ AlGaAs superlattice which
suppresses the tunneling of carriers. These dots are grown in a
single deposition of InAs at a substrate temperature of 470 $^{o}C$
and with a transition to self-assembled 3D growth at ~60 seconds.
The resulting dots are 2-3nm in height, and are capped in 5nm of
GaAs at 470 $^{o}C$ before raising the substrate temperature for
growth of the superlattice. $p$ and $n$ doping regions are arranged
symmetrically above and below the QD layer, with a total intrinsic
region thickness of 140 nm. The diode is encased in a weak planar
microcavity, with micron sized apertures in a metallic layer on the
surface to allow optical addressing of single dots.

\textbf{Experimental arrangement.} The sample is mounted inside the
bore of a superconducting magnet applying fields of up to 5T. When
the sample growth direction is aligned with the magnetic field a
single on-axis microscope objective is used to excite and collect
the emission from the sample. When the magnetic field is in the
plane of the sample (Voigt Geometry) an additional 45$^{o}$ mirror
is mounted to allow optical access to the sample. Photoluminescence
is excited from the sample with a continuous wave 850 nm laser
diode, and passed through a rotatable half-wave plate and polariser
before detection. For the data in Fig. \ref{Fig2} spectral
measurements confirm the sample was orientated within 0.1 $^{o}$ of
the magnetic field direction.

\parskip 5mm

\textbf{Acknowledgements}

\parskip 5mm

This work was partly supported by the EU through the Integrated
Project QESSENSE (project reference 248095), the Marie Curie Initial
Training Network (ITN) ``Spin-Optronics'' (project number 237252)
and EPSRC.

\textbf{Author Contributions} The samples were grown by I.F. and
D.A.R., and processed by M.A.P. The optical measurements were made
by A.J.B., M.A.P., Y.C. and N.S. A.J.S. guided the work. All authors
discussed the results and their interpretation. A.J.B. wrote the
manuscript, with contributions from the other authors.

\textbf{Competing Financial Interests}

The authors declare no competing financial interests.

\textbf{Supplementary Information}

\begin{figure}[ht]
\includegraphics[width=90mm]{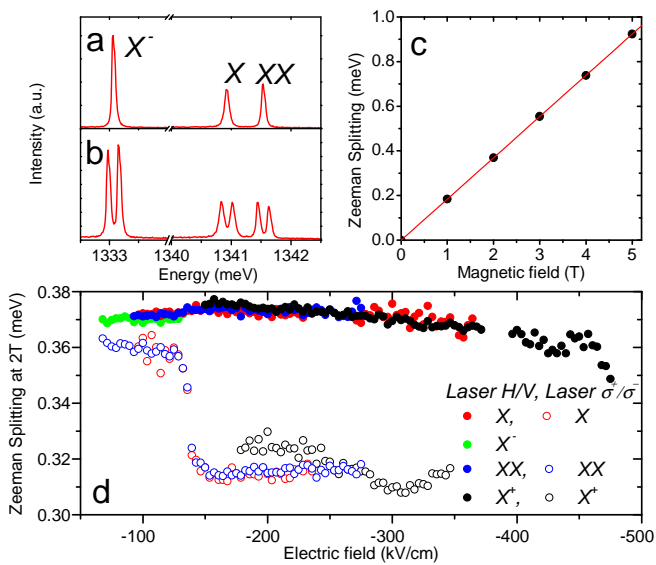}
\caption{\label{FigS1}: 
\textbf{Measurements   of  Faraday-geometry  $g$-factors.}  Emission
spectrum  at  (a) 0T and (b) 1T, both for an electric field of -57.1
kV/cm.  (c)  the splitting of the negatively-charged exciton,
$X^{-}$ at  -57.1  kV/cm,  as  the  magnetic field is changed. (d)
shows the Zeeman   splitting   of   $X^{-}$,   the   neutral exciton
($X$), neutral-biexciton  ($XX$)  and positively-charged exciton
($X^{+}$), as  a  function  of  electric  field.  The closed circles
indicate excitation  with  a  linearly  polarised laser, where  the
Zeeman splitting  is  proportional to the sum of the $s$-shell
electron and hole          $g$-factors, $|g^{\perp}_{\text{e,s}} +
g^{\perp}_{\text{h,s}}|$, and the open data points with a circularly
polarised  laser  sufficiently intense  to  partially polarise the
nuclear  field,  leading  to  a splitting  proportional to only the
$s$-shell      hole $g$-factor, $|g^{\perp}_{\text{h,s}}|$. }
\end{figure}

We also carried out measurements of the Zeeman splittings in the
Faraday geometry, where \textbf{$B^{\perp}$} is orientated out of
the sample plane. The 4 brightest $s$-shell transitions, neutral
exciton ($X$), biexciton ($XX$) and positively and negatively
charged exciton ($X^{+}$, $X^{-}$) are split into two components
with opposite circular polarisation. When excited by a linearly
polarised 850nm laser both circular polarisation components of each
transition are visible in the spectrum (Figure \ref{FigS1}b). The
splitting of each state is the same within error suggesting the
presence of differing numbers of carriers does not change the sum of
the Faraday $s$-shell electron and hole $g$-factors,
$|g^{\perp}_{\text{e,s}} + g^{\perp}_{\text{h,s}}|$. The Zeeman
splitting of $X^{-}$, shown in Figure \ref{FigS1}c, is proportional
to $|g^{\perp}_{\text{e,s}} + g^{\perp}_{\text{h,s}}| \mu_{\text{B}}
B^{\perp}$. There is a 3$\%$ change in this splitting from -150 to
-500 kV/cm (Figure \ref{FigS1}d, filled data points).

From the Zeeman splitting observed with linearly polarised
excitation we are only able to determine $|g^{\perp}_{\text{e,s}} +
g^{\perp}_{\text{h,s}}|$, not the separate $g$-factors. However,
optically pumping the system with a circularly polarised laser
preferentially creates one sign of electron spin in the quantum dot.
This imbalance of electron spin can partially polarise the nuclear
field through an electron-flip/nuclear-flop process, resulting in an
abrupt change in Zeeman splitting as the pump intensity is increased
\cite{Tarta2007}. Above this critical pump intensity the Zeeman
splitting is constant, because the nuclear field has been
sufficiently polarised to counteract the effect of the magnetic
external field. Now the Zeeman splitting is determined only by
$|g^{\perp}_{\text{h,s}}|$ so the two $g$-factors can be determined
separately. Measurements of the Zeeman splitting in the nuclear
pumping regime are shown in Figure \ref{FigS1}d as open data points,
with nuclear pumping only possible at fields less than -150 kV/cm.
Here the $X^{+}$ is dominant in the spectrum and it is the unpaired
electron in the upper state that is polarising the nuclear field.
Noise in the data means it is not possible to determine whether the
small variation in $|g^{\perp}_{\text{e,s}} + g^{\perp}_{h,s}|$ is
due to $|g^{\perp}_{\text{h,s}}|$ or $|g^{\perp}_{\text{e,s}}|$.
However, we infer that $|g^{\perp}_{\text{e,s}}| \sim$ 0.5 and
$|g^{\perp}_{\text{h,s}}| \sim $ 2.7, over a bias range of several
hundred kV/cm. These absolute values are consistent with previous
reports working at lower electric fields, which take the sign of
both $g$-factors to be negative.
\parskip 5mm


\end{document}